\documentstyle[aps,preprint,tighten,floats,epsf,rotate]{revtex}

\begin{document}
\draft
%%%%%%%%%%%%%%%%%%%%%%%%%%%%%%%%%%%%%%%%%%%%%%%%%%%%%%%%%%%%%%%%%%%%%%%%%

%\preprint{\vbox{\it 
%                    \null\hfill\rm IP-BBSR/2005-.., 4 Jan.'2005}\\\\}
%%%%%%%%%%%%%%%%%%%%%%%%%%%%%%%%%%%%%%%%%%%%%%%%%%%%%%%%%%%%%%%%%%%%%%%%%
%
\title{Excited hadrons as a signal for quark-gluon plasma formation}
\author{Biswanath Layek \footnote{e-mail: layek@iopb.res.in}, 
 Ajit M. Srivastava \footnote{e-mail: ajit@iopb.res.in}}
\address{Institute of Physics, Sachivalaya Marg, Bhubaneswar 751005, 
India}
\author{Soma Sanyal \footnote{e-mail: ssanyal@indiana.edu}}
\address{Department of Physics, Indiana University, Bloomington 
47405, USA}
%\date{Dec. 2004}
%
\maketitle
\widetext
\parshape=1 0.75in 5.5in
\begin{abstract}
  At the quark-hadron transition, when quarks get confined to hadrons,
certain orbitally excited states, namely those which have excitation 
energies above the respective $L = 0$ states of the same order as 
the transition temperature $T_c$, may form easily because of thermal 
velocities of quarks at the transition temperature. We propose that the 
ratio of multiplicities of such excited states to the respective $L = 0$
states can serve as an almost model independent signal for the 
quark-gluon plasma formation in relativistic heavy-ion collisions. For 
example, the ratio $R^*$ of multiplicities of 
$D_{SJ}^{*\pm}(2317)(J^P = 0^+)$ and $D_S^{*\pm}(2112)(J^P = 1^-)$ when 
plotted with respect to the center of mass energy of the collision 
$\sqrt{s}$ (or vs. centrality/number of participants), should show a 
jump at the value of $\sqrt{s}$ beyond which the QGP formation occurs.
This should happen irrespective of the shape of the overall plot
of $R^*$ vs. $\sqrt{s}$. Recent data from RHIC on $\Lambda^*/\Lambda$
vs. N$_{part}$ for large values of N$_{part}$ may be indicative of
such a behavior, though there are large error bars. We give a list of  
several other such  candidate hadronic states.
\end{abstract}
\vskip 0.125 in
\parshape=1 0.75in 5.5in
\pacs{PACS numbers: 25.75.-q, 12.38.Mh}
Keywords: quark-hadron transition, quark-gluon plasma, 
relativistic heavy-ion collisions
%\newpage
%\begin{multicols}{2}
\narrowtext
%%%%%%%%%%%%%%%%%%%

\section{Introduction}

There are strong reasons to  believe that in ultra-relativistic collisions
of heavy nuclei, a hot dense region of quark-gluon plasma (QGP) may get
created. There is a wealth of data which strongly suggests that
this transient QGP state may have been achieved in these experiments
\cite{qgprev}. Still a conclusive evidence for the QGP detection is 
missing. Among many signals which have been proposed for the detection
of QGP, probably the suppression of J/$\psi$ had the potential of being the 
cleanest signal \cite{jpsi}. However, due to various uncertainties involving
final state interactions etc. it has not been possible so far to use even 
this signal to make definitive statements about QGP detection \cite{jpsi2}.
Most other signals depend more sensitively on the details of plasma 
evolution as well as on the details of the quark-hadron transition dynamics, 
and hence the predictions become crucially dependent on underlying models 
used to describe the dynamics of various stages of the parton system 
evolution \cite{sgnl}.

 Ideally one would like to have a signal which depends least on the details 
such as the order of quark-hadron transition etc. In view of lattice 
results, which suggest that quark-hadron transition may be a 
cross-over \cite{lattice}, one would like to only use the single defining
property of the quark hadron transition, namely that quarks and gluons
are deconfined above the transition temperature $T_c$, while they are 
confined into hadronic states below $T_c$. (For a smooth cross-over, there 
may not be a sharp value of $T_c$ at which confinement properties of
quarks change abruptly. In such a case, by the transition temperature 
$T_c$ we will mean a range of temperature near the value $T = T_c$
across which the confinement of quarks significantly changes. If this
range is narrow enough then use of a definite value for $T_c$ will still 
capture the essential physics of our model.) In the following we argue that
the ratio $R^*$ of multiplicities of certain specific orbitally excited 
hadronic states to the respective $L = 0$ states, where $L$ is the 
orbital angular momentum, may provide such a model independent signal 
for QGP formation in relativistic heavy-ion collisions. (In the following, 
by orbitally excited state, and orbital excitation energy, we will always 
mean the $L = 1$ orbital excitation above the corresponding $L = 0$ state, 
and the energy splitting of these two levels.)  For example, the ratio 
$R^*$ of multiplicities of $D_{SJ}^{*\pm}(2317)(J^P = 0^+)$ and 
$D_S^{*\pm}(2112)(J^P = 1^-)$ when  plotted with respect to the center of 
mass energy of the collision $\sqrt{s}$, or centrality/number of participants 
(N$_{part}$),  should show a jump at the value $\sqrt{s_c}$ of $\sqrt{s}$ 
(or centrality) beyond which the QGP formation occurs. 

It is important to clarify here that we do not discuss
how $R^*$ should behave as a function of $\sqrt{s}$ (or, N$_{part}$). That
is a separate issue and involves factors like the final state interaction
etc. Our only claim is that, whatever be the form of the curve of $R^*$
vs. $\sqrt{s}$ (increasing or decreasing), 
there should be a jump enhancement at (or near the value) $\sqrt{s_c}$
(or, at some specific value of N$_{part}$) beyond which QGP formation
happens. In this context we note that the recent data from RHIC on
$\Lambda^*/\Lambda$ shows a strong decrease as a function of 
N$_{part}$ \cite{lambda1,lambda2}. However, as just mentioned above, 
this is of no consequence to our model. What is important is to look for 
any possible jump in this decreasing curve at some specific value of 
N$_{part}$. We note from  ref. \cite{lambda1} that the value of the 
multiplicity ratio for $\Lambda^*/\Lambda$ for very large N$_{part}$ 
(beyond 350) may be consistent with a slight increase, though completely 
within the error bar, to conclude anything. (This possible increase
is less clear in ref.\cite{lambda2}, which could be due to different
binning.) What is important is that if there is any rapid increase in $R^*$ 
beyond some N$_{part}$ then it will be very hard to explain it with existing
models where a decrease in the multiplicities of resonances
arising from final state interaction in the hadronic medium is naturally
expected. (Note that any resonance regeneration should show smooth
dependence on $\sqrt{s}$.) As we will see below, our model will naturally 
account for any such increase. 

We give a list of several hadronic states 
which could provide such a signal for QGP formation in relativistic heavy-ion
collisions.  We mention that in order that these enhanced multiplicities
of orbitally excited states survive up to the hadronic chemical freezeout, 
the chemical equilibrium should not last for long after the hadronization.
This may not be an unreasonable assumption as chemical freezeout
temperatures are close to the expected value 
of the transition temperature \cite{chemeq}. It is also expected that
resonances may not achieve chemical equilibrium. 

  The paper is organized in the following manner. In section II, we
describe the physical picture underlying the signal we are proposing.
In section III, we discuss the details of the signal and certain 
special features specific to this signal.  In section IV we discuss
a rough estimate of the ratio $R^*$ based on a recombination
model. In section V we give a list of candidate hadronic states 
which can be used for this signal. Section VI gives conclusions.

\section{Physical picture of the model}

   We start by trying to form a physical picture of the process of
hadron formation at the transition stage. We will restrict our attention
to hadrons with not very high $P_T$, so that the hadronization will
be expected to be dominated by something like the recombination 
process, rather than by fragmentation \cite{recomb,recomb2}. This may cover
hadrons starting from very low $P_T$, up to hadrons with $P_T$ as large  
as few GeV/c, see ref.  \cite{recomb}. (As we will later argue, there are 
other arguments in our model suggesting that the signal we propose will be
more prominent when restricted to low $P_T$ hadrons.) Further, as in certain 
recombination models \cite{recomb}, at the onset of hadronization, we may 
think of the effective degrees of freedom to be only (constituent) quarks 
with gluons essentially not being present as dynamical degrees of freedom. 
During hadronization, the effects of gluons will be primarily manifested in 
terms of confining color strings. In our picture the formation of hadrons
will be with valence quarks which seems to be a reasonable
approximation \cite{recomb}. 

 In the deconfined plasma, above $T_c$, the color field 
lines are spread out, and quarks and gluons are quasi-free with the 
usual Debye screening of color charges. As the transition temperature is
approached from above, confinement sets in and color field lines start
squeezing in the form of color flux tubes connecting quarks with antiquarks
forming mesons, or forming baryons, glueball etc. (We use here the language 
of color string for the simplicity of physical picture. Basic physics we 
use is simply that hadrons form by combination of nearby quarks/antiquarks, 
as in the recombination models  \cite{recomb,recomb2}).

For simplicity, let us first consider a single quark and antiquark with 
the overall quark-antiquark system being in the color singlet state. 
Let us further assume a hypothetical situation when the quark and the
antiquark do not have any kinetic energy as the system approaches the
transition stage. The color lines will squeeze in the form of a flux tube,
connecting the quark with the antiquark thereby forming a meson. If the
color field distribution about the line joining the quark and the antiquark 
was not very asymmetric then the resulting meson should most likely form in 
a state of zero orbital angular momentum. This is because the color field 
lines will just squeeze, forming the color flux tube, which will lead to the
confining potential for forming the meson. This flux tube, for a symmetric
configuration will not have any orbital angular momentum, but may have
radial excitations, as it relaxes to the appropriate meson state in
a given energy eigenstate. Even in a bag model like picture, initially 
static quark and antiquark will be attracted to each other, get bound
inside the bag with appropriate kinetic energies inside the bag,
developing the constituent masses. Again, for isolated pair, and
symmetric initial color fields, it seems unlikely that an orbital
excitation can result. 

   Orbitally excited states for this case also can form due to following 
types of processes. As the quark-antiquark pair, connected by the flux 
tube (or inside the bag, in the bag picture), relaxes to an appropriate 
meson state, it can create some other lighter meson by fragmentation 
(or emit soft gluons if the initial quark and antiquark
are not in a color singlet state, as in the Color Octet models
\cite{octet}), resulting in the generation of orbital angular momentum for 
the remaining meson.  Secondly, most often the initial color field 
configuration will be asymmetric (time dependent), leading to the 
formation of color flux tube (or bag) at the transition stage which is 
of deformed (time dependent) shape. Such a flux tube (bag), 
while relaxing to the meson state may generate orbital angular momentum. 
Also, orbital angular momentum can be generated due to the effects of 
other partons on this quark-antiquark pair.

  So far we have made the unrealistic assumption that the quark and the 
antiquark under consideration had no kinetic energies at the onset of 
hadronization. Let us now take into account the fact that at a non-zero 
temperature $T$,  quarks and antiquarks will have thermal energies, with 
average kinetic energy per degree of freedom being about $\frac{1}{2}T$. 
For relativistic quarks one should talk about the total energy. However, at
the onset of hadronization, quarks will develop constituent masses, so
even for light quarks, the use of non-relativistic terminology may not be
a serious problem, especially as our discussion is mostly qualitative in 
nature. Let us now consider the binding of such quarks and antiquarks into 
mesons at $T = T_c$. The flux tube  developing between the quark and the 
antiquark pair, now connects two point particles which have  non-zero
velocities. Clearly, in this case, for certain fraction (of order one) 
of cases, the quark and antiquark velocities will end up in the form
of non-zero angular momentum of the resulting hadronic state. For example
(using a semiclassical, non-relativistic picture), when their velocities are 
transverse to the flux tube connecting the pair, and are opposite to each 
other, then the resulting meson state will have non-zero orbital angular 
momentum. In the bag picture, with appropriate impact parameter, initial
velocities will lead to orbital excitation as quarks get confined into
hadrons. 

Note here that quarks will anyway develop appropriate kinetic 
energies as they fall into any bound state. Such kinetic energies will
result from the color forces binding the quarks into hadronic states.
As we mentioned above, this will happen even if quark/antiquark did not 
have any initial kinetic energies. However, if each quark and antiquark
in the entire system has a given (average) definite kinetic energy, then in 
certain fraction of the cases when these velocities are appropriately 
oriented, orbital excitation should result. What we are saying is that it is 
difficult to form $L = 0$ hadrons when quarks/antiquarks have large initial 
velocities, just as it is difficult to form orbitally excited
hadrons with initially static quarks and antiquarks. $L = 0$ meson can 
result only for special situations when the momenta of the quark and
the antiquark are either parallel, or antiparallel with zero impact
parameter. For generic situations, orbital excitation should result.   
(In the string picture, the string itself may have non-zero orbital angular 
momentum due to time dependence of the initial color field between the
quark and the antiquark, as we mentioned above. However, it is not clear
how, or whether, this string angular momentum may be related to thermal
energies of partons, gluons in this case. Presumably, there will be no
{\it net} orbital angular momentum left for the string if many gluons
participate during the development of the color flux tube between the 
quark and the antiquark at the onset of confinement. Any other 
contributions to string angular momentum should be present even when the 
system temperature remains below $T_c$, hence it should not affect the 
qualitative aspects of the signal we are proposing.)

  From the above intuitive reasoning, we conclude that thermal velocities
of quarks and antiquarks at the hadronization stage can lead to the 
formation of orbitally excited mesons. In this picture, the thermal kinetic
energies of the quark and the antiquark provide the required excitation
energy for creating the orbital excitation. Now, in general the kinetic
energy available will not be close to the required orbital excitation
energy. For example, suppose the separation between the $L = 0$ 
state and the orbitally excited state is of order 800 MeV or more 
(as for the orbital excitations of $\pi$ and $\eta$ mesons). Even the 
combined thermal kinetic energies of quark and the antiquark should not be 
much larger than about 2-3$T_c \simeq $ 500 MeV (with $T_c \simeq 170$ MeV).
In such a situation the required orbital angular momentum has to come from 
other sources as well, as discussed above.

  However if one considers those specific mesons for which the orbital
excitation energy roughly matches with the combined thermal kinetic energies 
of the quarks, then the required orbital excitation energy is right away
available for creating the excited meson state. As every quark/antiquark
has an average kinetic energy of order $T_c$ at the stage when they get
confined to hadrons, we conclude that formation of those orbitally
excited mesons which have orbital excitation energies of order $\alpha T_c$
should be greatly enhanced. Here the factor $\alpha$ is introduced to
parameterize the part of kinetic energy of the quarks that gets converted 
into the orbital excitation energy of the meson.  We are assuming here that
a significant part of thermal kinetic energy of quarks survives as quarks
develop constituent masses at the onset of confinement, even for light
quarks. We again mention that quarks, when they get bound into hadrons, 
will have appropriate kinetic energies (e.g. in the bag model), which will
result from confining color forces. What we are suggesting is that it
is possible that a part (characterized by $\alpha$) of initial 
{\it thermal} kinetic energies of quarks, when quark/antiquark velocities 
are appropriately matched, may end up directly in the form of orbital 
excitation.

With this, we can have a rough estimate of the range of $\alpha$ 
by first assuming that it is only one specific velocity component 
of, say, quark which will give the required orbital excitation energy, when 
combined with opposite velocity component of roughly similar magnitude 
of the partner antiquark. Each velocity component will have thermal kinetic 
energy of about $\frac{1}{2}T_c$ at the transition stage (again, using a 
non-relativistic picture with quarks which develop constituent masses). Thus 
the total thermal energy of the quark-antiquark system which can be used to 
give orbital excitation energy will be about $T_c$. Other velocity components 
of quark and the antiquark could combine to give higher orbital excitation 
energy, or could contribute to radial excitation energy. For example, 
average kinetic energy of the quark for the degree of freedom  transverse to
the color string will be about $T_c$. Combined kinetic energy of the
quark and antiquark for suitable configuration (with opposite velocities
for quark and antiquark) can then become 2$T_c$ which can directly
translate into the orbital excitation energy. From the above qualitative 
discussion we expect that the average value of $\alpha$ could be close 
to 1-2, and in any case should not be larger than 3. The value $\alpha = 3$
corresponds to the case when entire kinetic energy $\frac{3}{2}T$ of
the quark and of the antiquark ends up in the form of the orbital 
excitation energy. (of course, these are kinetic energies on the average
only.) For the case of baryons, the estimates of $\alpha$ should 
be appropriately modified by accounting for the thermal kinetic energy
of each of the three quarks.

\section{excited hadrons in relativistic heavy ion collisions}

  Let us now consider the general situation of a relativistic heavy-ion 
collision. Let us first consider the situation when the collision
process creates a system which thermalizes but does not reach upto
the transition temperature $T_c$ so the QGP state is never created. 
In such a situation the formation of hadrons  can be modeled using string
fragmentation as the QCD strings break creating secondary hadrons
\cite{frag}. In this situation there appears no reason to relate the thermal
energies (of hadrons) to the orbital excitation energies, at the
stage of hadron formation. The production of orbital excitations will 
mostly result from the transverse momentum created for the 
quark-antiquark pairs from the breaking of QCD string in the 
fragmentation model \cite{frag}. There will be a distribution of 
$P_T$ of quarks and antiquarks, with no specific value of $P_T$ preferred. 
In particular, there is no reason to relate the transverse momentum of 
quarks/antiquarks at the fragmentation stage to the thermal
energies in the resulting hadron system. Thus there is no reason
to expect enhancement in the multiplicities of {\it any specific} 
orbitally excited states at some particular value of the temperature of 
the final hadronic system.

  Now consider the situation when the temperature of the system reaches 
above $T_c$ and deconfined QGP state forms. As the QGP expands and cools,
its temperature drops to $T_c$ and partons combine to form hadrons.
As we argued above, now specific orbitally excited states, with
orbital excitation energies close to $\alpha T_c$ will form with
enhanced rate. We should note that the formation of hadrons in a QGP state
which is cooling through $T_c$ and the formation of hadrons in collisions 
where temperature never reaches upto $T_c$ may be of very different nature 
\cite{recomb,recomb2}. Therefore one may not be able to compare directly the
multiplicities of orbitally excited states between the two cases. However, 
we can consider the following ratio,

\begin{equation}
R^* \equiv {N(M^*) \over N(M)} .
\end{equation}

Here $N(M^*)$ and $N(M)$ are the total multiplicities of orbitally excited 
meson and the corresponding $L = 0$ state meson respectively.  For such a 
ratio, the effects of any possible intrinsic difference between the 
two types of hadronization may be minimized. Also, to make sure that
hadrons which form from hadronization of QGP do not get mixed with
hadrons forming in the regions which do not undergo deconfining
transition one should confine to near central rapidity regions. As
we have mentioned above, restriction to hadrons with not very high
$P_T$ should ensure that only hadrons forming via a recombination
type process are being included and not the ones coming from
the fragmentation process. 

 As for other orbitally excited states, for which the orbital excitation
energy is very different from the expected range of $\alpha T_c$, for them 
also the ratio $R^*$ defined above may be somewhat enhanced due to thermal 
velocities of quarks and antiquarks. However, for them the dominant 
contribution to the orbital excitation energy must come from sources other 
than the thermal energy. Those sources will lead to contributions which may
become difficult to distinguish from various sources for the case of
hadronization when temperature remains below $T_c$. Hence, in those cases 
the ratio $R^*$ may not show any significant change as the maximum
temperature of the system crosses $T_c$.

  It is not entirely clear what happens for the states which have excitation
energies well below $T_c$. Though, in the particle data table there appear 
to be no such states, still, conceptually one should  discuss this case.
If thermal kinetic energy of quarks and antiquarks is very large then
during the formation of hadrons the remaining kinetic energy has to be
either converted to part of the energy of QCD string, or dissipated away
in other modes, e.g. by emission of lighter mesons.

 The effects we have discussed above are valid for baryons as well, as the 
basic physics of orbital excitation coming from the transverse components of 
thermal velocities of quarks remains the same. It is also possible that 
the effect we are proposing here may be present even for radially excited 
states of mesons. However, radial excitations may not be uncommon even when 
quarks do not have significant thermal velocities. For example, when a quark and 
an antiquark become joined by the QCD string at the stage of hadronization, 
then while relaxing to an appropriate meson state, radial excitations can 
result.  It may thus become difficult to disentangle the effect of thermal
kinetic energies on radial excitations. We therefore focus on the
orbital excitations of hadrons as the underlying physical picture is
simple to analyze. 

  The physical picture which we are using here for the enhanced formation
of specific orbitally excited states is similar to the formation of
resonances where the cross-section shows a peak when the center of mass 
energy equals the energy of the intermediate bound state. The difference 
here is that we are here focusing primarily on the initial kinetic energy 
of the quark and antiquark and essentially arguing that the cross-section 
should show enhancement when the initial kinetic energy is of the same order
as the orbital excitation energy of the meson state. Clearly this special
treatment of initial thermal kinetic energies of quarks is a very strong 
assumption, even stronger than the non-relativistic, semiclassical picture 
underlying the entire discussion.  For heavy quarks the non-relativistic 
treatment is reasonably accurate \cite{heavyq}. Even for light quarks, 
as the quarks start binding into hadrons the constituent 
quark mass will start playing the dominant role 
and the non-relativistic treatment may still capture essential physics.
(Velocities of light quarks are about 0.7-0.8 in heavy-light bound states 
\cite{lightq}.) What remains a gross approximation is the special 
status of the initial (thermal) kinetic energies of quarks/antiquarks.
As we discussed above, quarks and antiquarks, when they get bound 
into hadrons will develop kinetic energies (e.g. in the bag model), 
which will result from confining color forces. What we have focused
on is that the system undergoing hadronization consists of quarks
and antiquarks (as we mentioned earlier, we consider only on the valence
quarks/antiquarks in hadron formation), all of which have roughly
same kinetic energies (of order $T_c$). When such rapidly moving quarks 
and antiquarks get captured into hadrons, then it seems very natural 
to assume that it will be relatively difficult to form $L = 0$ hadrons 
and that a part (characterized by $\alpha$) of initial {\it thermal} kinetic 
energies of quarks may end up directly in the form of orbital excitation.

 As confinement sets in, and 
quarks pick up constituent masses, the kinetic energies of quarks will
contribute to resulting potential energy of the system, e.g.
to the energy of flux tube which connects a quark and an antiquark
to form a meson. Kinetic energies of quarks will also get significantly
affected as they develop constituent masses due to effects of confinement.
Such effects should tend to lower the value of $\alpha$, and clearly all
such effects have to be properly accounted for. For example, the 
spin-orbit interaction energy also may play an important role in 
determining the value of $\alpha$. In the absence of a detailed picture 
of confinement, we just parameterize such effects in terms of $\alpha$. 

  Even with all these ambiguities and crudeness of our arguments, we
suggest that there remains the possibility that the 
thermal velocities of quarks and antiquarks may have a sizable 
contribution to the orbital excitation of hadrons (and possibly also
for radial excitations of hadrons). If this is true then the relative 
ratios of orbitally excited  states $R^*$ (as discussed above) carry a
signature of the transition temperature $T_c$ at which all the 
quarks/antiquarks become bound into hadrons.

  The specific signal we propose is the following. One should plot the 
ratio $R^*$ for various hadronic states as a function of the center 
of mass energy of the collision, for, say, given colliding nuclei (which 
determines the initial energy density, and hence the temperature of the 
resulting system \cite{temp}). Alternatively, one can also plot $R^*$ as 
a function of number of participants N$_{part}$/centrality 
for a given center of mass energy. For generic 
hadronic states, $R^*$ should vary in a certain smooth manner as a 
function of $\sqrt{s}$ (or, of centrality). For example, 
if we assume that abundances of hadrons are determined by the thermal 
equilibrium distribution, as seems to be the case for most hadrons 
\cite{chemeq,unvrsl}, then the dominant behavior of  the ratio $R^*$
should be given by, $R^*_{thermal} \sim exp[-(E^*-E)/T]$ (with appropriate 
chemical freezeout temperature).
Here $E^*$ and $E$ are the energies of the orbitally excited hadron and 
the $L = 0$ state hadron respectively. We are not writing here the detailed
expressions for the multiplicities (including possible factors of
the appropriate chemical potentials), as in the thermal models \cite{chemeq}.
Also, we do not discuss about the contributions of decays of heavier states
to various multiplicities at freezeout.
This is because such factors will be present for the ratio $R^*$ even when
the system temperature always remains below $T_c$, and hence do not affect 
the nature of the signal we propose. For our purpose it suffices that
$R^*$ is some smooth function of $\sqrt{s}$. 

  It is important to clarify here that we are mentioning this case of
thermal abundances {\it only} as an example. We are not arguing for any 
specific shape of
the plot of $R^*$ vs. $\sqrt{s}$ (or, N$_{part}$). For example, it is
very likely that resonances do not chemically equilibrate, so the assumption of
thermal distribution for them is inaccurate. Further, the final multiplicities
will be severely modified by factors such as final state interactions and
loss of signals in the medium, which  will be expected to strongly suppress 
resonances as a function of $\sqrt{s}$ (or centrality), 
(also by enhancement factors such as resonance regeneration). 
The overall curve of $R^*$ may very well be decreasing as a function of 
$\sqrt{s}$ or N$_{part}$ (as for $\Lambda^*(1520)/\Lambda$, see ref. 
\cite{lambda1,lambda2}). 

  Now, at some critical value $\sqrt{s_c}$ 
of $\sqrt{s}$ at which the temperature of the system
reaches above $T_c$, the above described enhancement of orbitally
excited states should become operative due to the formation of
QGP state. This should result in an abrupt enhancement in the value of 
$R^*$ for those hadrons for which the orbital excitation energy
is close to $\alpha T$ at $T = T_c$, i.e.

\begin{equation}
E^* - E \simeq \alpha T_c .
\end{equation}

 Further, this enhancement should prevail at all higher values of 
$\sqrt{s}$. Again, note that if the overall curve of $R^*$ is decreasing
as a function of $\sqrt{s}$ then our argument implies that beyond 
a critical value $\sqrt{s_c}$ (or centrality), there should be a 
separate segment of the plot of $R^*$ with a jump enhancement at 
$\sqrt{s_c}$.  Even for other hadrons for which the orbital excitation 
energy is not too close to $\alpha T_c$, there may be some enhancement 
in the corresponding $R^*$ beyond $\sqrt{s_c}$, with the enhancement 
decreasing as $E^* - E$ deviates more and more from the value of 
$\alpha T_c$. One can also consider the ratio of multiplicity of an 
orbitally excited hadron state, say with energy $E^*$, to $L = 0$ state 
of some different hadron which has the same (or almost the same) energy. 
Such a ratio, from thermal models, should not strongly depend on temperature
(assuming other factors like chemical potential to be same for
the two states), and hence on $\sqrt{s}$, apart from the effects of
final state interaction in the hadronic medium. Our model will 
then predict a sharp change in the value of this ratio at some 
specific value $\sqrt{s_c}$ of $\sqrt{s}$. This may have an advantage that 
departure from a (possibly) roughly constant value of the ratio at 
$\sqrt{s_c}$ may be easier to identify. On the other hand, using states
of the same hadron is advantageous in minimizing the effects of details of
the parton system evolution etc.

  Note that this requires that the enhanced multiplicities of these specific
orbitally excited states should depart from the thermal distribution. 
This appears to be the case for resonances. In any case this 
possibility does not seem unreasonable to us as, even with the general
success of statistical models, it is clearly a very  strong 
assumption  that {\it strictly all} hadronic multiplicities should 
follow equilibrium  distributions. For this to happen, the hadronic system 
should last for long enough time (compared to all relevant hadronic 
interaction rates) before the chemical freezeout happens. This does not seem 
to be the case as chemical freezeout temperatures do not seem to be much 
different from the expected value of the transition temperature \cite{chemeq}. 
In fact, it has been argued in ref. \cite{chemeq} that, in view of the short 
time scale between the hadronization and the chemical freezeout, hadronic 
multiplicities are most likely established during hadronization. As the 
detailed dynamics of this process is not understood, is not clear to us how 
(or whether) this can lead to a suppression of the signal we are proposing. 
It is possible that if there is a large enough contribution to the 
multiplicities of certain specific orbitally excited states, as we are 
proposing, then the resulting jump in $R^*$ may survive.

   We do not know the value of $\alpha$ so one cannot say exactly where
this enhancement should occur. Still at some value of $T$, and hence of 
$\sqrt{s}$ (or centrality), the 
enhancement in the value of $R^*$ as described above
should occur. Such an abrupt enhancement will signal the formation of QGP, 
and at the same time, the value of $T$ (extracted from the value of 
$\sqrt{s}$) \cite{temp} will give the value of the transition temperature 
$T_c$ if we can control the estimate of $\alpha$. As we argued above we 
expect $\alpha$ to be roughly in the range of 1-3, with more probable values
being close to 1-2. Thus the value of $\alpha T_c$ can be conservatively 
estimated to lie somewhere in the range of 150 - 500 MeV, with a value of
about 150-350 MeV being a more probable estimate.

 We emphasize here three important aspects of the signal we have proposed. 
First, a jump in the values of $R^*$ at a common value $\sqrt{s_c}$ of 
$\sqrt{s}$ (or of centrality), for a set of hadrons, 
will be a concrete signal that the system
has gone through a transient stage of QGP phase. Secondly, from the value of
$\sqrt{s_c}$ one can extract the value of $T_c$ depending on the value
of $\alpha$, as mentioned above. This determination of $T_c$, in some
sense, will be similar in character to the determination of $T_c$ through the
value of $\sqrt{s}$ at which the proposed $J/\psi$ suppression \cite{jpsi}
takes place due to QGP formation. (Though, note that $J/\psi$ suppression
happens at a significantly higher temperature than $T_c$, above about 2$T_c$, 
see ref. \cite{jpsi3}. Also, $J/\psi$ suppression is not expected to be
sharp at some definite $\sqrt{s}$, it gets smeared because of final state 
interactions. As there appears no reason to expect such strong smearing in
our model, the jump in $R^*$ may be sharper.) Apart from the uncertainties 
in $\alpha$, this determination of $T_c$ will 
also depend on the details assumed for the entire collision 
process which relate the center of mass energy of the collision to the 
energy density of the plasma, and hence temperature by assuming thermal
equilibrium \cite{temp}. Various parameters, such as the initial 
thermalization time, crucially affect these estimates of the temperature.
The third feature of our proposed
signal is important from this point of view, as here one will have another 
completely independent determination of the approximate value of $T_c$.
This arises from the fact that the specific orbitally excited states
of hadrons which will show the jump in corresponding values of $R^*$
are precisely those which have orbital excitation energies of order 
$\alpha T_c$. For example, the candidate states we have proposed (see 
below) have been chosen because the expected value of $T_c$ is about 170 
MeV. (From this point of view, one should plot $R^*$ vs. $\sqrt{s}$ for 
all available partner hadronic states, to find out which ones, if any, 
show a jump in $R^*$.) Of course, as the value of $\alpha$ is uncertain 
at least by a factor of 2, one will only be able to infer the value
of $T_c$ within that factor by looking at the orbital excitation energies 
of states which show jump in $R^*$. Still, it is important to appreciate
that this determination of $T_c$ is completely independent of the 
determination of $T_c$ through $\sqrt{s_c}$ (or specific centrality)
{\it at which} the jump in $R^*$ is expected to occur, and is uncertain 
only by factor $\alpha$. (Of course, for better estimates of $\alpha$ one 
needs a proper understanding of the hadronization process itself.)
    
  We  mention another important effect which naturally happens in our model. 
In our picture, the thermal kinetic energies of quarks get used up
in generating the orbital excitation energies for certain specific
hadrons (as given above). Thus, for these hadrons, immediately
after formation, there should be little thermal kinetic energy left
for the center of mass motion of the hadron. For other hadrons for which
the thermal kinetic energies of the constituent quarks does not result 
in orbital excitation, net kinetic energy of hadron should be significant.
Thus, one expects that orbitally excited states should have low $P_T$,
while the $L = 0$ state hadrons should have $P_T$ as consistent with the
thermal kinetic energies involved. We will use this fact in the estimation
of $R^*$ using a recombination model in the next section. 
As the kinetic energy of the order of 
$\alpha T_c$ gets used up in creating the orbital excitation, we expect
that, on the average, $P_T$ of the excited states should be determined
using kinetic energies which are lower by about 
$\alpha T_c$ compared to other states.  Of course, this difference exists
only at the stage of hadronization. If thermal equilibrium of hadrons
prevails for any significant time after hadronization, all hadrons will
acquire thermal kinetic energies. However, for certain collisions
it is possible that hadronic thermal freezeout happens almost immediately 
after hadronization \cite{freeze}.  (For example, with large value of 
$\sqrt{s}$ to allow for QGP formation, but with small nuclei, so that 
transverse expansion becomes dominant at an early stage.) If that happens 
then orbitally excited states (or, in general even the radially excited
states) should have significantly low $P_T$ compared to other hadrons.
(In principle, this effect may be operative for even those orbitally
excited states which have significantly larger excitation energies
compared to $\alpha T_c$, as part of the orbital excitation energy may still
be coming from the thermal kinetic energies of quarks.) These arguments
have the important implication that for such collisions (where thermal
freezeout happens close to hadronization), the jump in 
$R^*$ which we have proposed to occur at some critical value of $\sqrt{s}$, 
should be much more prominent if one restricts to hadrons with low $P_T$.

\section{estimate of $R^*$ in recombination model}

 Some estimate of the ratio $R^*$ can be obtained within the framework of
recombination model discussed in ref. \cite{recomb}. For rough
estimates, we first review the non-relativistic estimate of 
the momentum distribution of the ground state meson $M$ from
ref.\cite{recomb}. Though we have quoted data for $\Lambda$ baryon from
RHIC (\cite{lambda1,lambda2}), we will not attempt to calculate $R^*$
for baryons in the recombination model to keep the basic physical arguments
simple. After all, recombination models do not capture the physics of
development of confining forces (string) between quarks and antiquarks.
Therefore, the calculations provided in this section still
have to be supplemented with the basic physics of our model, namely,
certain situations will be argued to lead to $L= 0$ meson while others
leading to $L = 1$ meson.
 
Spatial wave functions for a (quasi) free quark and antiquark 
(with momenta ${\bf p_1}$ and ${\bf p_2}$), and for a meson (with
momentum ${\bf P}$) formed by this quark-antiquark system are written as,

\begin{eqnarray}
{\nonumber} <x|Q,{\bf p_1,p_2}> \sim e^{i({\bf p_1.x_1} + {\bf p_2.x_2})} ,\\
<x|M,{\bf P}> \sim e^{i{\bf P.R}} \phi_M({\bf y}) .
\end{eqnarray}

 Here, ${\bf R} = ({\bf x_1} + {\bf x_2})/2$ is the center of mass coordinate 
and ${\bf y} = {\bf x_1} - {\bf x_2}$ is the relative coordinate. 
$\phi_M({\bf y})$ is the normalized meson wave function. The relative 
momentum conjugate to ${\bf y}$ is ${\bf q} = ({\bf p_1} - {\bf p_2})/2$, 
and the center of mass momentum is ${\bf P} = {\bf p_1} + {\bf p_2}$.

 Meson momentum distribution  is then evaluated  using the 
overlap $<Q,{\bf p_1},{\bf p_2}|M,{\bf P}>$ 
(see, ref. \cite{recomb} for details). One gets,

\begin{equation}
{dN_M \over d^3P} = A \int {d^3q \over (2\pi)^3} w({{\bf P} \over 2} + {\bf q})
w({{\bf P} \over 2} - {\bf q}) |{\hat \phi}_M({\bf q})|^2 .
\end{equation}

Here, $w({\bf q})$ gives  the parton phase space distribution, 
${\hat \phi}_M({\bf q})$ is the meson wave function in momentum space, and
$A$ accounts for constant factors like degeneracy and volume etc.

   Let us consider the situation when the momenta ${\bf p_1}$ and ${\bf p_2}$ 
of the quark and the antiquark are large (compared to the width of
${\hat \phi}_M({\bf q})$, which will be expected to be of order 
$\Lambda_{QCD}$). We then consider two situations,

(1) ${\bf p_1}$ and ${\bf p_2}$ are almost parallel, 
as discussed in ref.\cite{recomb}.  Then $|{\bf q}| << |{\bf P}|$.
As we discussed, this situation will correspond to the formation of $L = 0$ 
meson, with the meson center of mass momentum ${\bf P}$ being large. 

(2) ${\bf p_1}$ and ${\bf p_2}$ are almost opposite to each other. Then 
${\bf P}$ is small and $|{\bf q}| >> |{\bf P}|$. This is the opposite limit 
of the recombination case discussed in ref. \cite{recomb}, where this 
situation is not considered as leading to formation of any meson. For example,
in ref.\cite{recomb}, it is argued that hadrons with large ${\bf P}$ can
result from combining quarks with smaller momenta. We take this to hold true
for $L = 0$ meson, as mentioned in the case (1) above. However, for $L = 1$
hadrons we take the opposite situation where quarks with large momenta
combine to give a hadron with small ${\bf P}$. In ref. \cite{recomb},
higher $L$ states are incorporated by using
degeneracy factors, as in the statistical models. In contrast,  
in our model, almost opposite momenta of quark/antiquark
result in an orbitally excited meson, (with appropriate impact
parameter between the quark and the antiquark). Note that with long range
confining forces operating between quarks/antiquarks during hadronization,
combining even fast moving quark and antiquark (with opposite velocities)
to form a meson should not be a problem. We can then incorporate
this basic difference to modify the approach of ref. \cite{recomb} for
discussing $L = 1$ meson. 

  We first repeat the main steps of case (1) from ref.\cite{recomb}, and
then modify those steps for case (2).  For case(1), the meson multiplicity
is calculated (\cite{recomb}) by using the Taylor expansion of 
$w({{\bf P} \over 2}  \pm {\bf q})$ about
${{\bf P} \over 2}$ (as $|{\bf q}| << |{\bf P}|$). This gives,

\begin{equation}
w({{\bf P} \over 2} + {\bf q}) w({{\bf P} \over 2} - {\bf q}) = 
w({{\bf P} \over 2})^2 + O({\bf q}^2) .
\end{equation}

  With this, the lowest order contribution to the meson multiplicity
can be obtained (\cite{recomb}) just by using the normalization of the
wave function of the meson. One gets,

\begin{equation}
{dN_M \over d^3P} = A w({{\bf P} \over 2})^2 = A e^{-{{\bf P}^2 \over 4mT}} .
\end{equation}

 Here we have used the non-relativistic distribution for the quarks with mass
m (for simplicity we use same flavor for quark and the antiquark),

\begin{equation}
w({\bf P}) \sim e^{-{{\bf P}^2 \over 2mT}} .
\end{equation}

  Important thing to note in Eq.(6) is that to this order the expression for
the meson distribution does not depend on the shape of the meson wave 
function, which, for the ground state meson can be taken as
(in momentum space)

\begin{equation}
{\hat \phi}_M({\bf q}) \sim e^{-{\bf q}^2/2\Lambda_M^2} .
\end{equation}

$\Lambda_M$ is the width of the wave function, and should be of the order
of $\Lambda_{QCD}$. 

Now we consider case(2) and discuss the formation of $L = 1$ meson.
Since for this case we consider $|{\bf P}| << |{\bf q}|$, 
we Taylor expand $w({{\bf P} \over 2} \pm {\bf q})$
about $\pm {\bf q}$. We get,

\begin{equation}
w({\bf q} + {{\bf P} \over 2}) w(-{\bf q} + {{\bf P} \over 2}) = 
w({\bf q}) w(-{\bf q}) + O({\bf P}) .
\end{equation}

 Neglecting $O({\bf P})$ terms, we get,

\begin{equation}
{dN_{M^*} \over d^3P} = A' \int {d^3q \over (2\pi)^3} w({\bf q})
w(-{\bf q}) |{\hat \phi}_{M^*}({\bf q})|^2 .
\end{equation}

 $A'$ differs from $A$ in Eq.(4) in the degeneracy factor. It is clear 
that in this limit ($|{\bf P}| << |{\bf q}|$), the form of the wave function
will be important in determining the momentum distribution of the meson
$M^*$. Since in this case we are considering formation of $L = 1$ 
meson, we use the following form of the wave function for 
meson $M^*$ (\cite{hrmn}),

\begin{equation}
{\hat \phi}_{M^*}({\bf q}) \sim q_x e^{-{{\bf q}^2 \over 2 \Lambda_M^2}} .
\end{equation}

Note that we use the same width $\Lambda_M$ as for $L = 0$ meson.  Here, $q_x$ 
is the x component of ${\bf q}$. Using Eq.(7) for $w(\pm {\bf q})$, we get,

\begin{equation}
{dN_{M^*} \over d^3P} = A' \int {d^3q \over (2\pi)^3} 
e^{-{{\bf q}^2 \over mT}} q_x^2 e^{-{{\bf q}^2 \over \Lambda_M^2}} .
\end{equation}

 Let us define,

\begin{equation}
{\bf q}' = {\bf q}(1 + {\Lambda_M^2 \over mT})^{1/2} \equiv {\bf q}  C .
\end{equation}

 Then we can write,

\begin{equation}
{dN_{M^*} \over d^3P} = {A' \over C^5} \int {d^3q' \over (2\pi)^3}
{q'_x}^2 e^{-{{{\bf q}'^2} \over \Lambda_M^2}} .
\end{equation}

 The integral is unity due to the normalization of the wavefunction. We 
thus get,

\begin{equation}
{dN_{M^*} \over d^3P} = {A' \over C^5} .
\end{equation}

 Using this equation and Eq.(6) we get the following expression for
the (momentum dependent) ratio $R_p^*$.

\begin{equation}
R_P^* = { \kappa e^{{\bf P}^2 \over 4mT} \over (1 + \Lambda_M^2/mT)^{5/2}} ,
\end{equation}

 with $T$ here being equal to $T_c$.
Here, $\kappa$ gives the ratio of the two degeneracy factors.
This expression shows a rapid increase for large value of ($L = 0$) meson 
momentum. One may consider this increase as coming from the trivial 
factor of $e^{-{\bf P}^2/2mT}$ for the momentum distribution of the $L = 0$
meson, as the $L = 1$ meson is taken here to have small momenta. However,
it is important to note here that both mesons ($L = 0$ and $L = 1$) are
considered here as forming from the same quark and antiquark having
large momenta ${\bf p_1}$ and ${\bf p_2}$. There is a suppression factor 
for both, quark, and antiquark, of
the form $e^{-{{\bf p_i}^2/2m}}, ~ i=1,2$. For $L = 0$ meson, this leads
to large momenta ${\bf P}$ for the meson, leading to the suppression factor
$e^{-{{\bf P}^2/2m}}$. On the other hand, for $L = 1$ meson, the same 
suppression  factors of the quark and the antiquark are absorbed in the 
wave function  of the meson. This softens the suppression factor from 
an exponential suppression to a multiplicative factor as in Eq.(16). 
Note that this expression should eventually breakdown at sufficiently high
${\bf P}$. This is because our assumption that a quark and an antiquark
of opposite momenta leads to an (orbitally excited)  bound state 
will not hold true for very large values of
quark/antiquark momenta. Even for $L = 0$ meson, as discussed in 
ref. \cite{recomb}, recombination model is not expected to be applicable
for very large values of ${\bf P}$ (beyond a couple of GeV), as the
physics may be dominated by fragmentation in this regime. 

 We should clarify that we are not claiming here a derivation showing the
enhancement of $R^*$ as we have argued in previous sections. For example,
one could have very well considered the $L = 0$ meson wave function 
for case (2), instead of Eq.(11), with similar result as in Eq.(16)
with the exponent 5/2 changing to 3/2. (Or, one could have considered other
excited states of the meson.) The above estimate of $R_P^*$ 
has to be supplemented with our basic picture that for a quark and an 
antiquark with large momenta ${\bf p_1}$ and ${\bf p_2}$, 
$L = 0$ meson is likely to result when the two momenta are almost parallel. 
On the other hand, if the  two momenta are almost opposite to each other, 
then $L = 1$ meson (or, possibly a radially excited one which we do not 
discuss) is likely to form because the available thermal kinetic energy may
be enough to lead to $L = 1$ excitation (at least for the candidate states
we suggest in the next section). The estimate of $R^*$ provided here is, 
therefore, to be taken as indicative of the extra contribution to the 
multiplicity of the orbitally excited meson $M^*$ when QGP formation occurs.    

\section{candidate hadronic states for the signal}

  Using the particle data table \cite{data}, there are several candidate 
hadrons which have orbitally excited states with excitation energies of 
order 200-400 MeV above the corresponding $L = 0$ states. We give several 
such candidates in the following. Other candidate states can be found
from ref. \cite{data} with the basic criterion being the relevant range
for the orbital excitation energy. Below, each equation gives the hadronic 
state, along with its partner orbitally excited state.

\vskip .3in

{\bf Mesons :}

\begin{eqnarray}
{\nonumber} D_S^{*\pm} (J^P = 1^-) ~ m = 2112 ~{\rm MeV} \\ 
{\nonumber} D_{SJ}^{*\pm} (J^P = 0^+) ~ m = 2317 ~{\rm MeV} \\
\Delta m = 205 ~ {\rm MeV}
\end{eqnarray}

\begin{eqnarray}
{\nonumber}D^{*0} (J^P = 1^-) ~ m = 2007 ~{\rm MeV} \\
{\nonumber}D_2^{*0} (J^P = 2^+) ~ m = 2459 ~{\rm MeV} \\
\Delta m = 452 ~{\rm MeV}
\end{eqnarray}

\begin{eqnarray}
{\nonumber}J/\psi(1S) (J^P = 1^-) ~ m = 3097 ~{\rm MeV} \\
{\nonumber}\chi_{C0}(1P) (J^P = 0^+) ~ m = 3415 ~{\rm MeV} \\
\Delta m = 318 ~{\rm MeV}
\end{eqnarray}

\vskip .3in
{\bf Baryons :}

\begin{eqnarray}
{\nonumber}\Lambda_C^+ (J^P = \frac{1}{2}^+) ~ m = 2285 ~{\rm MeV} \\
{\nonumber}\Lambda_C^+ (J^P = \frac{1}{2}^-) ~ m = 2594 ~{\rm MeV} \\
{\nonumber}\Lambda_C^+ (J^P = \frac{3}{2}^-) ~ m = 2625 ~{\rm MeV} \\
\Delta m = 309, 340~{\rm MeV}
\end{eqnarray}

\begin{eqnarray}
{\nonumber}\Lambda (J^P = \frac{1}{2}^+) ~ m = 1116 ~{\rm MeV} \\
{\nonumber}\Lambda (J^P = \frac{1}{2}^-) ~ m = 1406 ~{\rm MeV} \\
{\nonumber}\Lambda (J^P = \frac{3}{2}^-) ~ m = 1519 ~{\rm MeV} \\
\Delta m = 290, 403~{\rm MeV}
\end{eqnarray}

\begin{eqnarray}
{\nonumber}\Xi_C^+ (J^P = \frac{1}{2}^+) ~ m = 2466 ~{\rm MeV} \\
{\nonumber}\Xi_C^+ (J^P = \frac{3}{2}^+) ~ m = 2647 ~{\rm MeV} \\
{\nonumber}\Xi_C^+ (J^P = \frac{1}{2}^-) ~ m = 2790 ~{\rm MeV} \\
\Delta m = 324, 143 ~{\rm MeV}
\end{eqnarray}

\begin{eqnarray}
{\nonumber}\Xi_C^0 (J^P = \frac{1}{2}^+) ~ m = 2472 ~{\rm MeV} \\
{\nonumber}\Xi_C^0 (J^P = \frac{3}{2}^+) ~ m = 2645 ~{\rm MeV} \\
{\nonumber}\Xi_C^0 (J^P = \frac{1}{2}^-) ~ m = 2790 ~{\rm MeV} \\
\Delta m = 318, 145 ~{\rm MeV}
\end{eqnarray}

 We mention that for some of these states $J^P$ as quoted above are not 
confirmed, see ref.\cite{data}. The energy differences for the partner
states as given above are reasonably close to the expected value of 
$\alpha T_c$ (with $T_c \simeq 170$ MeV), and hence should be good candidates 
for evaluation of the ratio $R^*$ as described above. For example, 
$D_{SJ}^{*\pm}(2317)(J^P = 0^+)$ and $D_S^{*\pm}(2112)(J^P = 1^-)$ 
have a mass difference of 205 MeV and hence one may expect that an important 
fraction of the orbital excitation energy in this case may come from the 
thermal kinetic energies of quark and antiquark (even with lower estimates 
of $\alpha$ as mentioned above). In principle one could allow the spin of 
the $L = 0$ state and the corresponding orbitally excited 
state to be different. In view of hyperfine contributions 
one may like to have the same spin for the partner states for the 
evaluation of the ratio $R^*$. However, such contributions 
will be present for the formation of these states even below $T_c$, and for 
hadrons forming at $T_c$ in our picture it may just affect the value of 
$\alpha$ without affecting the qualitative aspect of the signal (i.e., a
jump in the value of $R^*$ at $\sqrt{s_c}$). We have not addressed here the
important issue of experimental observations of all these states mentioned 
above, and for many of these it may be difficult to have data in
relativistic heavy-ion collisions. 

 Recently, measurements of $\Lambda^*(1520)/\Lambda$ have become available
from RHIC as a function of N$_{part}$ \cite{lambda1,lambda2}. Note that this 
set is one of the candidates we have listed above (Eq.(21), with $\Lambda^*$ 
listed at 1519 MeV). The multiplicity ratio $R^*$ for  $\Lambda^*/\Lambda$ 
shows a strong decrease as a function of N$_{part}$ \cite{lambda1,lambda2}. 
As discussed above, this overall decrease is of no consequence to our
model. However, the value of the multiplicity ratio for very large N$_{part}$ 
(beyond 350) may be consistent with a slight increase, though completely within 
the error bar, to conclude anything definite.
We again stress, what is important is that if there is any increase 
in $R^*$ beyond some N$_{part}$ then it will be very hard to explain it 
with existing models where a rapid decrease in the multiplicities of 
resonances arising from final state interaction in the hadronic medium is 
naturally expected. Our model will naturally account for any such increase. 
In fact, the (possible) increase in $R^*$ seems to occur beyond
sufficiently large value of N$_{part} \simeq 350$ which may be consistent
with QGP formation. We mention that this (possible) rise in $R^*$ for
$\Lambda^*/\Lambda$ is more distinct in ref.\cite{lambda1} where
$R^*$ is plotted as a function of N$_{part}$. In ref. \cite{lambda2}, the
plot is w.r.t. $dN_{ch}/d\eta$, and the rise does not appear clear (which
may be due to different binning). One has to wait for more refined data 
with a more detailed plot of $R^*$ vs. N$_{part}$ to see if actually there 
is a discontinuity in the plot of $R^*$ for $\Lambda^*(1520)/\Lambda$ 
beyond some value of N$_{part}$.

  It is important to realize here that the abundances of the hadrons may 
be dependent on QGP formation from other physical considerations also.
For example, the suppression of $J/\psi$ and $\chi_{c_i}$ due to 
Debye screening of color in the QGP state is well discussed in the 
literature.  Further, due to larger radius of $\chi_{c_i}$, the 
suppression from Debye screening is larger for $\chi_{c_i}$ \cite{jpsi3}. 
Even in the recombination model, one would expect that forming orbitally
excited states may be harder due to their larger 
sizes as the medium is very dense.
Clearly such effects will give rise to opposite behavior for the ratio $R^*$ 
as compared to the effect of thermal kinetic energy of quarks as we have 
discussed. Final dependence of $R^*$ on $\sqrt{s}$ will result from a 
combination of the two effects.  (Note that similar effects also occur from
hadronic final state interaction. However, such {\it hadronic state
interaction effects} are of no consequence to our model which only predicts a 
jump in $R^*$ beyond some $\sqrt{s}$, and {\it not} the overall shape of
the curve of $R^*$.
Thus, decrease of $\Lambda^*/\Lambda$ vs. N$_{part}$ may be naturally
expected from such final state interactions, but any possible rapid rise 
beyond some value of N$_{part}$ cannot be accommodated in such models.)
 Though there are detailed estimates \cite{jpsi3} of suppression 
of $J/\psi$ and $\chi_{c_i}$, it is not possible to say which effect will 
dominate due to lack of definite quantitative estimates for the 
effect we are proposing. In any case, for hadrons with light quarks, the Debye 
screening is very effective due to larger sizes of these hadrons (compared 
to the Debye length), so one only needs to consider their formation from 
quark/antiquark recombinations at $T_c$. Thus, for these hadrons, the
predicted jump in  $R^*$ should not be adversely affected by these other
considerations.

\section{conclusions}

   In conclusion we have suggested a novel signature for the detection
of the transient QGP state formation in relativistic heavy-ion collisions.
The basic nature of the signal is almost model independent in the sense that
it does not depend on the nature or even the existence of phase transition
between the confined and the deconfined state of QCD. Further, details
of plasma evolution etc. also may not be of much importance. This is 
because $R^*$ depends on the ratios of multiplicities of states of a given 
hadron and presumably the effects of plasma evolution etc. will get
minimized in taking the ratio. 

   We emphasize that though the arguments we have presented are very 
qualitative and crude, the basic physics underlying these arguments is
simple. Main argument we have used is that quarks/antiquarks have thermal 
velocities at the stage when they all get captured to form bound states at 
the hadronization stage, i.e. when the temperature of the system is equal 
to $T_c$. Due to this thermal velocity, when QCD string starts
joining, say, a quark with antiquark to form a meson, it is easy to
generate orbital motion of quarks in the bound states. Essential part of
the argument being that it is difficult to form $L = 0$ hadrons when 
quarks/antiquarks have large initial velocities, in the same way as it 
is difficult to form orbitally excited hadrons if quarks/antiquarks
did not have any initial velocities.  We then further argue that for those 
hadrons for which the orbital excitation energy is a certain fraction
(characterized by $\alpha$) of the thermal kinetic energy of quark/antiquark 
system at $T = T_c$, the formation of such orbitally excited bound states 
will be enhanced.  Using these arguments we suggest that the ratio of 
multiplicities of orbitally excited state to the $L = 0$ state of such specific 
hadrons should show an abrupt increase at a critical value of $\sqrt{s}$ which 
corresponds to the formation of an intermediate, transient, QGP state. 
As we mentioned above, alternatively, one can also plot $R^*$ as a function 
of centrality for a given center of mass energy, with $R^*$ expected
to show a jump above certain value of the number of participants.

The signal we have proposed is simple and just requires plotting of the 
respective ratios $R^*$ (as defined by Eq.(1)) for various hadrons with 
respect to $\sqrt{s}$. One important feature of this signal is that it 
predicts a jump in the value of $R^*$, at a common value of $\sqrt{s}$ (for 
given colliding nuclei), for a whole class of hadronic states which have
orbital excitation energy splitting of similar magnitude. Most likely candidate 
hadrons which have orbital excitation energies roughly in the required range 
are charmed and strange hadrons. For 
example, $D_{SJ}^{*\pm}(2317)(J^P = 0^+)$ and 
$D_S^{*\pm}(2112)(J^P = 1^-)$ have a mass difference of 205 MeV. Thus, a plot 
of the ratio of total multiplicities of $D_{SJ}^{*\pm}(2317)(J^P = 0^+)$ 
and $D_S^{*\pm}(2112)(J^P = 1^-)$ with increasing $\sqrt{s}$ may contain 
the information of QGP formation as discussed above. We have given a list 
of several such candidate hadronic states. We note that the recent data
\cite{lambda1,lambda2} on the multiplicity ratio $\Lambda^*(1520)/\Lambda$ 
from RHIC as a function of N$_{part}$ may be consistent with our model, with a 
possible hint of a slight increase beyond N$_{part} \simeq 350$. 
One needs to have a more refined data, with more detailed plot for
intermediate values of N$_{part}$ to make any definite conclusions.
We have not addressed the issue  of experimental observations for various
resonances listed in the previous section. It may not be easy to collect enough 
statistics for many of these candidate states \cite{openc,jpsi2}. 
As we mentioned above, the effect we have proposed may be
present even for radially excited states of hadrons. However, for
radial excitations the effect may not be prominent. We have primarily 
discussed orbital excitations of hadrons as the basic physics of our 
arguments is simpler to model for this case. A better understanding of
the underlying physics of quark/antiquark recombination into hadrons
is needed before one can make any definitive quantitative predictions
regarding this signal.

\vskip .2in
\centerline {\bf ACKNOWLEDGEMENTS}
\vskip .1in

 We are very grateful to S.A. Bass, D.K. Srivastava, J. Cleymans, and
S. Digal for very useful discussions and comments. We also thank S.C. 
Phatak, B. Rai, A. P. Mishra, R. Ray, and V. S. Kumar for useful discussions. 

%%%%%%%%%%%%%%%%%%% 

%\end{multicols}

\begin{thebibliography}{99}

\bibitem{qgprev} See, for example, U. Heinz, hep-ph/0407360. 

\bibitem{jpsi} T. Matsui and H. Satz, Phys. Lett. {\bf B 178}, 416 (1986).

\bibitem{jpsi2} See, for example, M. Brooks, J. Phys. G {\bf 30}, 
S861 (2004).

\bibitem{sgnl} D. Zschiesche et al., Heavy Ion Phys. {\bf 14},
425 (2001).

\bibitem{lattice} Z. Fodor and S.D. Katz, Prog. Theor. Phys. Suppl. 
{\bf 153}, 86 (2004); A. Peikert, F. Karsch, E. Laermann, B. Sturm,
Nucl. Phys. Proc. Suppl. {\bf 73}, 468 (1999).

\bibitem{lambda1} L. Gaudichet, J. Phys. G. {\bf 30}, S549 (2004);
C. Markert, nucl-ex/0308029.

\bibitem{lambda2} C. Markert, nucl-ex/0501033; P. Fachini, 
nucl-ex/0403026.

\bibitem{chemeq} J. Cleymans and K. Redlich, 
Phys. Rev. {\bf C 60}, 054908 (1999), K. Redlich, J. Cleymans,
H. Oeschler, and A. Tounsi, Act. Phys. Pol. {\bf B 33}, 1609 (2002).

\bibitem{recomb} R.J. Fries, B. Muller, C. Nonaka, and
S.A. Bass, nucl-th/0306027.

\bibitem{recomb2} K.P. Das and R.C. Hwa, Phys. Lett. 
{\bf B 68}, 459 (1977).

\bibitem{octet} E. Braaten and S. Fleming,  Phys. Rev. Lett. 
{\bf 74}, 3327 (1995)

\bibitem{frag} H. Bengtsson and T. Sjostrand, Comp. Phys.
Comm. {\bf 46}, 43 (1987); A. Casher, H. Neuberger, and 
S. Nussinov, Phys. Rev. {\bf D 20}, 179 (1979).

\bibitem{heavyq} G.S. Bali, Phys. Rep. {\bf 343}, 1 (2001).

\bibitem{lightq} D. Ebert, V.O. Galkin, and R.N. Faustov,
Phys. Rev. {\bf D 57}, 5663 (1998).

\bibitem{temp} K.J. Eskola, K. Kajantie, P.V. Ruuskanen, and
K. Tuominen, Nucl. Phys. {\bf B570} (2000) 379; K.J. Eskola, 
P.V. Ruuskanen, S.S. Rasanen, and K. Tuominen, Nucl. Phys.
{\bf A 696}, 715 (2001).

\bibitem{unvrsl} F. Becattini, hep-ph/9701275, Z. Phys. {\bf C 69}, 
485 (1996).

\bibitem{jpsi3} S. Datta, F. Karsch, P. Petreczky, and I. Wetzorke,
J. Phys. {\bf G30}, S1347 (2004); P. Petreczky, S. Datta, F. Karsch, 
and I. Wetzorke, Nucl. Phys. Proc. Suppl. {\bf 129}, 596 (2004),  
M. Asakawa and T. Hatsuda, J. Phys. {\bf G30}, S1337 (2004).

\bibitem{hrmn} D. Faiman and A.W. Hendry, Phys. Rev. 
{\bf 173}, 1720 (1968).
 
\bibitem{data} S. Eidelman et al. (Particle Data Group), Phys.
Lett. {\bf B 592}, 1 (2004).

\bibitem{freeze} S.A. Bass and A. Dumitru, Phys. Rev. {\bf C61},
064909 (2000); A. Dumitru, Phys. Lett. {\bf B463}, 138 (1999);
T. Csorgo and L.P. Csernai, Phys. Lett. {\bf B333}, 494
(1994); J.P. Bondorf, H. Feldmeier, I.N. Mishustin, and
G. Neergaard, Phys. Rev. {\bf C65}, 017601 (2002).

\bibitem{openc} Z. Xu, nucl-ex/0410005.

\end{thebibliography}
\end{document}